\documentclass[12pt,preprint]{emulateapj}
\usepackage{amssymb}
\usepackage{times}
\usepackage{ulem} % for the strikethrough
\usepackage{amsmath}
\usepackage{multirow}
\usepackage[colorlinks,backref,dvipdfm,linkcolor=blue,anchorcolor=,citecolor=blue]{hyperref}
\makeatletter
\newcommand{\teff}{$T_{\mathrm{eff}}$}
\newcommand{\muhz}{$\mu$Hz}
\newcommand{\numax}{$\nu_{\mathrm{max}}$}
\newcommand{\dnu}{$\Delta\nu$}
\newcommand{\kepler}{\textit{Kepler}}

\newcommand{\kic}{\textit{KIC}}
\newcommand{\keplermission}{\textit{Kepler Mission}}
\newcommand{\eps}{$\epsilon$}
\newcommand{\msol}{M$_\odot$}

\newcommand{\tosc}{$T_{\mathrm{osc}}$}
\newcommand{\rmnum}[1]{\romannumeral #1}
\newcommand{\Rmnum}[1]{\expandafter\@slowromancap\romannumeral #1@}
\makeatother

\slugcomment{}

\shorttitle{Asteroseismic study of RGB stars in cluster}
\shortauthors{Wu et al.}

\begin{document}


%\title{Asteroseismic study of red-giant stars in NGC 6791 and NGC 6819 using \textit{Kepler} data}
\title{New asteroseismic scaling relations based on Hayashi track relation applied to red-giant branch stars in NGC 6791 and NGC 6819}


\author{
T. Wu\altaffilmark{1,2,3}, Y. Li \altaffilmark{1,2}, S. Hekker\altaffilmark{4,5}
}


\altaffiltext{1}{Yunnan Observatories, Chinese Academy of Sciences, P.O. Box 110, Kunming 650011, P. R. China; wutao@ynao.ac.cn, ly@ynao.ac.cn, hekker@mps.mpg.de}
\altaffiltext{2}{Key Laboratory for Structure and Evolution of Celestial Objects, Chinese Academy of Sciences, P.O. Box 110, Kunming 650011, P. R. China}
\altaffiltext{3}{University of Chinese Academy of Sciences, Beijing 100039, P. R. China}
\altaffiltext{4}{Max Planck Institute for Solar System Research, Max Planck Strasse 2, 37191 Katlenburg-Lindau, Germany}
\altaffiltext{5}{Astronomical institute 'Anton Pannekoek', University of Amsterdam, Science Park 904, 1098 XH, Amsterdam, the Netherlands}


\begin{abstract}
 Stellar mass $M$, radius $R$, and gravity $g$ are important basic parameters in stellar physics. Accurate values for these parameters can be obtained from the gravitational interaction between stars in multiple systems or from asteroseismology. Stars in a cluster are thought to be formed coevally from the same interstellar cloud of gas and dust. The cluster members are therefore expected to have some properties in common. These common properties strengthen our ability to constrain stellar models and asteroseismically derived $M$, $R$ and $g$ when tested against an ensemble of cluster stars.
 Here we derive new scaling relations based on a relation for stars on the Hayashi track ($\sqrt{T_{\rm eff}} \sim g^pR^q$) to determine the masses and metallicities of red giant branch stars in open clusters NGC 6791 and NGC 6819 from the global oscillation parameters  \dnu\ (the large frequency separation) and \numax\ (frequency of maximum oscillation power). The \dnu\ and \numax\ values are derived from \kepler\ observations.
 From the analysis of these new relations we derive:
(1) direct observational evidence that the masses of red giant branch stars in a cluster are the same within their uncertainties,
(2) new methods to derive $M$ and $z$ of the cluster in a self consistent way from \dnu\ and \numax, with lower intrinsic uncertainties,
(3) the mass dependence in the \dnu\ - \numax\ relation for red giant branch stars.
\end{abstract}



\section{Introduction}\label{sec1}
%Asteroseismology:
Asteroseismology is a powerful tool to obtain different kinds of detailed information about the internal structure and evolutionary state of stars which cannot be obtained from classical (non-timeseries) data. It is possible to determine fundamental stellar parameters (mass, radius, surface gravity and mean density) from the observed oscillation parameters as these are defined by the stellar internal structure. Thanks to the space-based observations from instruments, such as WIRE \citep{Hacking99,Buzasi00}, MOST \citep[][]{Walker03-MOSTmission}, CoRoT \citep{Baglin06}, and \kepler\ \citep{Koch10,Gilliland10}, more and more stars with solar-like oscillations have been observed. This has opened a new research area in which it is possible to study large samples of stars, i.e., so-called ``ensemble asteroseismology" \citep{Chaplin11}.

%solar-like oscillation:
Solar-like oscillations are stochastically excited by the convective turbulence in the stellar envelope \citep[e.g.][]{Frandsen02} and are expected to be present in all stars with convective outer layers, such as low-mass main-sequence (MS) stars, subgiants, stars on the red-giant branch (RGB), as well as red giants in the helium core burning phase on the horizontal branch or red clump (RC) and asymptotic-giant branch (AGB).

%cluster:
Stars in a cluster are thought to be formed coevally from the same interstellar cloud of gas and dust. The cluster members are therefore expected to have some properties in common. These restrictions strengthen our ability to constrain stellar models when tested against an ensemble of cluster stars \citep[e.g.][]{Audard96,Hekker11b,basu11,Miglio12}. Therefore, asteroseismology of clusters is a potentially powerful tool for improving our understanding of stellar evolution.

%cluster research history:
Prior to \kepler, many attempts were undertaken to detect solar-like oscillations in open and globular clusters \citep[see][for a summary]{Stello10b}. \citet{Gilliland93} aimed to detect oscillations in turn-off stars in the open cluster M67 during a dedicated multi-site campaign. In this study, although an impressively low level of noise was obtained, the stellar oscillations could not be detected. However, there was a red-giant star in the observed field that showed the evidence of excess power in the excepted frequency range. Unfortunately, individual modes of the star could not be resolved from the data due to the limited length of the time series (about one week). Inspired by the above work, \citet{Stello07a,Stello07b} aimed at detecting oscillations in red-giant stars in M67 during a 6-week long multi-site campaign. Strong evidence for excess power in the Fourier spectra was found for a number of stars, but they were not able to clearly disentangle the noise and oscillation signal in the analysis of those stars. At the same time, several attempts to detect stellar oscillations in globular clusters were carried out. \citet{Frandsen07a,Frandsen07b} studied the red giants in globular cluster M4 and provided lower limits on oscillation amplitudes, indicating that the low metallicity stars of M4 might have lower oscillation amplitudes than the empirically predicted value of $A\propto L/M$ \citep{kjeldsen95}. \citet{Edmonds-and-Gilliland96} and \citet{Stello_Gilliland09} observed the globular clusters 47 Tuc and NGC 6397, respectively, using the Hubble Space Telescope (HST). \citet{Stello_Gilliland09} aimed at detecting solar-like oscillations in a sample of red giants in the metal-poor globular cluster NGC 6397 and used those results to test the scaling relations for the mode amplitudes. In their results, only one star showed evidence of oscillations, but there was no unambiguous detection of the solar-like oscillation or of equally spaced frequencies in this star.

%the newest cluster asteroseismology research status:
\citet{Stello10a} obtained the first clear detections of solar-like oscillations in red giants in the open cluster NGC 6819 from observations with \kepler. They were able to measure the large frequency separation, \dnu, and the frequency of maximum oscillation power, \numax. Soon afterwards, a number of other results were presented including determinations of cluster membership \citep{Stello10a,Stello11b}, investigations of the oscillation-amplitude relation \citep{Stello11a}, determinations of scaling relation for global oscillation parameters \dnu\ and \numax\ \citep{Hekker11b}, measurements of RGB masses and radii and cluster distance modulus and age \citep{basu11}, investigations of cluster RGB mass loss properties from hydrogen-shell to core-helium burning phases \citep{Miglio12}, and tests of small frequency separations between modes of degree $0$ and $2$  and the phase term $\epsilon$ \citep[see Eq.\ref{eq_dnu1}][]{Corsaro12}.

% the present study:
In this paper, we devise a new method to estimate the masses and metallicities of RGB stars in  NGC 6819 and NGC 6791 from global oscillation parameters (\dnu\ and \numax).

\section{Solar-like oscillations: \dnu\ and \numax}\label{sec-solar-like-oscillation}
According to the asymptotic theory \citep{Vandakurov67,Tassoul80,Gough86}, frequencies of solar-like oscillations are regularly spaced and approximately expressed as follows:
\begin{equation}\label{eq_dnu1}
\nu_{n,l}\approx\Delta\nu(n+\frac{1}{2}l+\epsilon),
\end{equation}
where $n$ is the radial order and $l$ is the spherical harmonic degree. The quantity \dnu\ is the large frequency separation of oscillation modes with the same degree and consecutive order, i.e. $\Delta\nu=\nu_{n,l}-\nu_{n-1,l}$. \dnu\ is approximately the inverse of twice the sound travel time from the stellar center to the stellar surface and proportional to the square root of the mean density of the star \citep[for more details see e.g.][]{kjeldsen95}:
\begin{displaymath}
\Delta\nu\approx(2\int_{0}^{R}\frac{dr}{c_{\rm{s}}})^{-1}\propto\sqrt{\bar{\rho}},
\end{displaymath}
where $c_{\rm{s}}$ is acoustic velocity and $\bar{\rho}$ is the mean density of the star. The large frequency separation is to second order a function of frequency $\nu$ and the degree $l$. In practice, often a mean value $\langle\Delta\nu\rangle$ over a specified frequency range is used to characterize a star. The quantity \eps\ is a phase-term.

The large frequency separation \dnu\ scales with mass $M$ and radius $R$ as described by \citet{kjeldsen95}:
\begin{equation}\label{eq_deltanu}
\Delta\nu=\sqrt{\frac{M/M_{\odot}}{(R/R_{\odot})^3}}\Delta\nu_{\odot},
\end{equation}
where $\Delta\nu_{\odot}=134.88~\mu$Hz is taken from \citet{Kallinger10}.

The oscillations are centered at the so-called frequency of maximum oscillation power, which is denoted as \numax. At this frequency the pulsation amplitude reaches its maximum. \numax\ scales with $M$, $R$ and $T_{\rm eff}$ as follows:
\begin{equation}\label{eq_numax}
\nu_{\mathrm{max}}=\frac{M/M_{\odot}}{(R/R_{\odot})^2\sqrt{T_{\mathrm{eff}}/T_{\mathrm{eff},\odot}}}\nu_{\mathrm{max},\odot}
\end{equation}
\citep{Brown91,kjeldsen95,Belkacem11}. $T_{\rm{eff},\odot}$ = 5777 K and $\nu_{\rm{max},\odot}$ = 3120 $\mu$Hz are taken from \citet{Kallinger10}.

The scaling relations are successfully used for main-sequence stars, subgiants \citep{Bedding-kje03}, and red-giant stars \citep[e.g.][]{Stello08,Kallinger10}, although for the latter with increased uncertainties \citep[e.g.][]{White11,Miglio12,Mosser13,hekker13}.

\section{Analyses of observational data}\label{sec3}

\subsection{Target selection}
We investigate two open clusters observed by the NASA \kepler\ satellite, i.e. NGC 6791 and NGC 6819. The cluster membership and stellar evolutionary phases were identified by \citet{Stello11b}. We exclude the non-members and red-clump stars, and only select the red-giant branch stars. We subsequently remove stars for which no unambiguous \dnu\ could be determined. Therefore, the target selection is based on three criteria: \rmnum{1}. they are members of the cluster; \rmnum{2}. they are red-giant branch stars; \rmnum{3}. they have unambiguous  detections of \dnu. This results in 42 and 31 targets for NGC 6791 and NGC 6819, respectively (see Fig. \ref{fig_cmd}).

\subsection{Fitting isochrones and estimating effective temperatures}\label{sec-iso}

It is commonly assumed that all stars in a cluster are formed simultaneously from the same cloud of interstellar dust and gas. This means that stars in a cluster have homogeneous properties (including age, distance, composition, etc.), except for the stellar masses. Here, we do not consider merging of clusters nor regeneration of stars in a cluster. The most commonly used method of obtaining cluster information is to fit isochrones, which we also do.

We use the Modules for Experiments in Stellar Astrophysics (MESA) evolutionary code, which is developed by \citet{mesa}. The theoretical isochrones of a cluster are computed using the ``multimass" model of MESA. Based on the default parameters, we adopt the OPAL opacity GS98 series, and treat the convection zone by the standard mixing-length theory (MLT) with the mixing-length parameter $\alpha_{\mathrm{MLT}} = 1.95$ and consider the convective core overshooting with $f_{\rm{ov}} = 0.06$. The stellar atmosphere model is described by \citet{ks66}. These authors defined the effective temperature at optical depth $\tau_{\rm{s}} = 0.312$, i.e. $T_{\rm{eff}}=T(\tau_{\rm{s}}=0.312)$. For the detailed description of the model parameters see \citet{mesa}. Element diffusion, semi-convection, thermohaline mixing, mass loss, and rotation are not included in the present study.

\begin{figure}
\begin{center}
\includegraphics[scale=0.6,angle=-90]{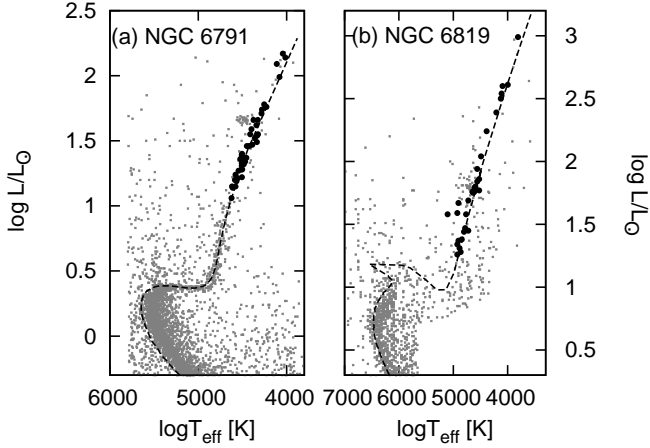}
\caption{Herzsprung Russell diagram. Panel (a) and (b) show cluster NGC 6791 and NGC 6819, respectively. In the two panels, the dark grey points indicate cluster members; the targets investigated in this study are shown by the large black dots; the dashed lines represent isochrones (see Sect. \ref{sec-iso}); information of these isochrone is provided in Table~\ref{table_p-color}.}\label{fig_cmd}
\end{center}
\end{figure}

The $B$ and $V$ photometric data for NGC 6791 and NGC 6819 are from \citet{Stetson03} and \citet{Hole09}, respectively. Following \citet{Hekker11b}, we adopt a value of 0.01 mag as random uncertainty in each color band for all stars. This results in an uncertainty of about 0.02 mag in the color index $(\bv)$.
To obtain the cluster parameters, we transform $(\bv)$ into \teff, using the color-temperature calibrations by \citet{rm05}, and transform $V$ into $\log L$, using the bolometric correction relation established by \citet{Flower96} and the corrected and modified coefficients by \citet{Torres10}. We fit the isochrones in the HR diagram (see Fig. \ref{fig_cmd}). For NGC 6791 we obtain a metallicity $z=0.040\pm0.005$, reddening $E(\bv)=0.14\pm0.01$ mag, distance modulus $(m-M)_{V}=13.36\pm0.04$ mag, and age $\tau_{\rm{age}}=8.0$ Gyr. For NGC 6819, we obtain a metallicity $z=0.022\pm0.004$, reddening $E(\bv)=0.13\pm0.01$ mag, distance modulus $(m-M)_{V}=12.40\pm0.05$ mag, and age $\tau_{\rm{age}}=1.85$ Gyr (see Table~\ref{table_p-color}). For NGC 6791, the distance modulus $(m-M)_{V}$ is slightly lower than $13.51\pm0.09$ mag determined by \citet{Brogaard12} from multiple eclipsing binary systems. For NGC 6819 the distance modulus is consistent with $12.37\pm0.10$ mag determined by \citet{Jeffries13} from a detached eclipsing binary system.

Using the RM05 color-\teff\ relations to estimate \teff, we find that for giants an uncertainty of 0.01 in the color index $(\bv)$ leads to an uncertainty of 40 K in \teff, while an uncertainty of 0.1 dex in [Fe/H] leads to an uncertainty of 20 K in \teff. Additionally an uncertainty of 0.01 in $E(\bv)$ leads to an uncertainty of 20 K in \teff, and finally the system uncertainty is altogether 50 K. This results in a total uncertainty in \teff\ of  40 K (color index) + 20 K (reddening) + 20 K (metallicity) + 50 K (system) = 130 K. For the field stars, we use the effective temperatures from the \kepler\ Input Catalog \citep[KIC][]{Brown11}, with $\sim$150 K total uncertainty.

\begin{deluxetable*}{lcccccc}
\tablewidth{0pt}
\tablecaption{Information of isochrones.\label{table_p-color}}
\tablehead{
\colhead{ } & \colhead{ } & \colhead{$E(\bv)$} & \colhead{$(m-M)_{{{V}}}$} & \colhead{$M_{\rm{turnoff}}$} & \colhead{$M_{\mathrm{RGB}}$} & \colhead{Age} \\
\colhead{Clusters} & \colhead{z} & \colhead{[mag]} & \colhead{[mag]} & \colhead{[\msol]} & \colhead{[\msol]} & \colhead{[Gyr]}
}
\startdata
NGC 6791  & 0.040$\pm$0.005 & 0.14$\pm$0.01 & 13.36$\pm$0.04 & 1.09 & 1.15 & 8.0 \\
NGC 6819  & 0.022$\pm$0.004 & 0.13$\pm$0.01 & 12.40$\pm$0.05 & 1.68 & 1.74 & 1.85
\enddata
\end{deluxetable*}

\subsection{Analyses of power spectra}\label{sec-analysis}

To compute power spectra we use the \kepler\ lightcurves provided by the \kepler\ Asteroseismic Science Consortium. The time series of photometric data were obtained between 2009 and 2012 (Q0--Q13, while some of the target stars lack one or two observational quarters). They consist of approximately 43000 data points per star obtained in the LC (Long-Cadence) observational mode with a cadence of about half an hour ($\Delta t=29.4$ minutes). The data are corrected for instrumental effects following \citet{garcia11}. We transform the light curves to relative flux, with unit ppm, i.e.:
\begin{displaymath}
fl(t)=(\frac{f(t)}{\bar{f}}-1)\times10^{6},
\end{displaymath}
where $f(t)$ is the corrected flux of a star, and $\bar{f}$ is its mean flux. We compute the Fourier transformation for the relative flux $fl(t)$ to obtain a power spectrum.

\subsubsection{Background and \numax}\label{sec-fit-background}

\begin{figure*}
\begin{center}
\includegraphics[scale=0.67,angle=-90]{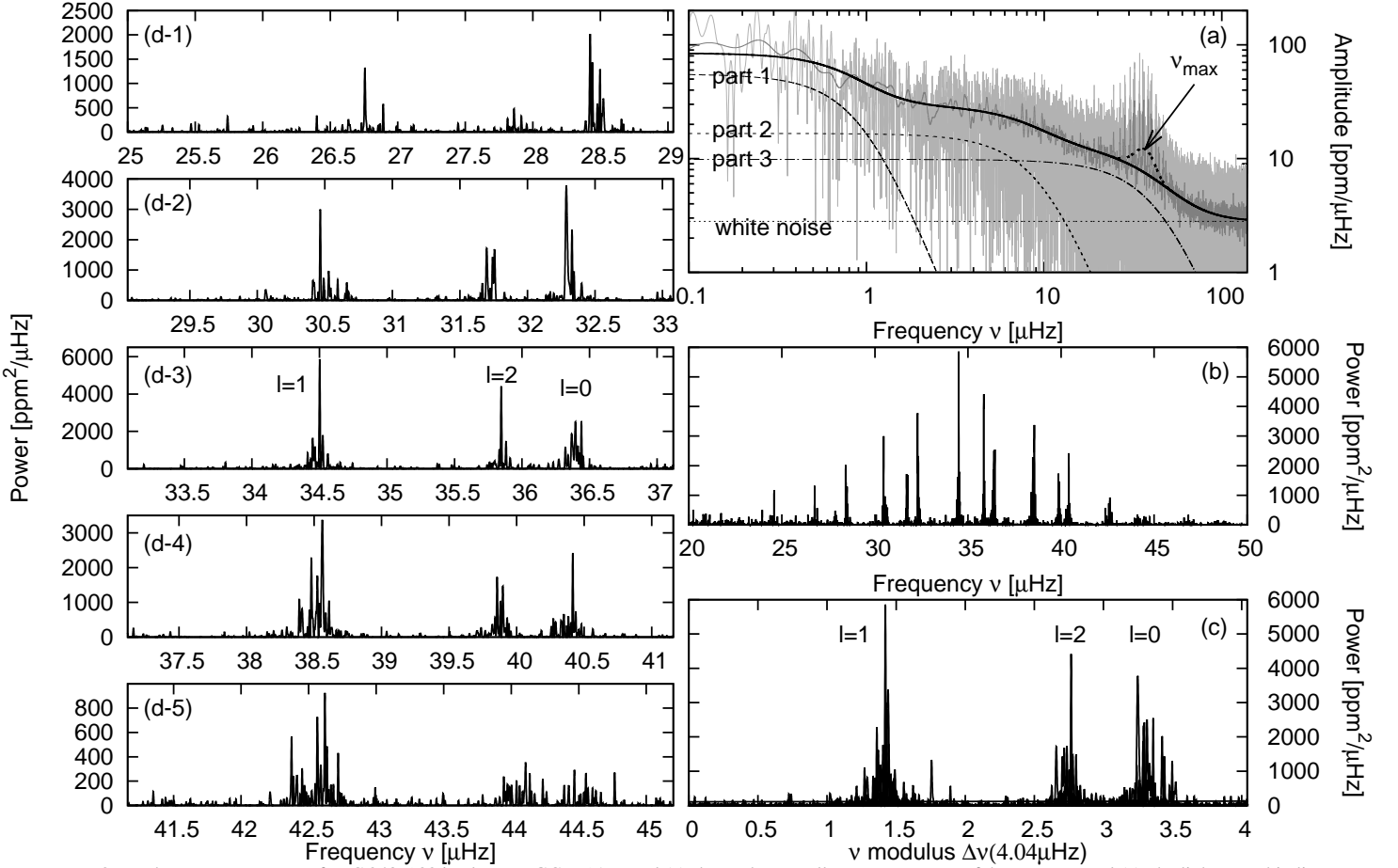}
\caption{The power spectrum of KIC 2436900 in cluster NGC 6791. Panel (a) shows the overall power spectrum of the star. In panel (a), the light grey thin line represents the original power spectrum; the dark grey thin line shows the smoothed power spectrum; the black dotted line is the white noise; the dashed-dotted line, short-dashed line, and the long-dashed line are the frequency-dependent background signals, respectively, which are corresponding to the Lorentzian-like terms of Equation \eqref{backg}; the thick solid line shows the total background noise and the thick dotted line represents the fit of the total power spectrum. Panel (b) shows the power spectrum in the frequency range of the oscillations. Panel (c) shows an alternative form of an \'{e}chelle diagram. The left five panels, panels (d-1)--(d-5), show the different orders of the observed oscillations. The degrees $l$ are presented in panels (c) and (d-3).}
\label{power}
\end{center}
\end{figure*}

The power spectra are composed of a background signal (i.e.~background noise) and pulsations. To extract the pulsation information, we remove the background signal from the power spectrum. The background signal consists of frequency-dependent signals due to, for example, stellar activity, granulation, and faculae as well as frequency-independent white noise. These frequency-dependent signals can be modeled by a sum of several Lorentzian-like functions  \citep{Harvey85}:
\begin{equation}\label{eq-bg-Harvey}
P_{\rm{stellar}}(\nu)=\frac{4\sigma^2\tau}{1+(2\pi\nu\tau)^2},
\end{equation}
or represented by modified Lorentzian-like functions:
\begin{equation}\label{eq-bg-Karoff}
P_{\rm{stellar}}(\nu)=\frac{4\sigma^2\tau}{1+(2\pi\nu\tau)^2+(2\pi\nu\tau)^4}.
\end{equation}
The latter was introduced by \citet{Karoff08}. For different components of the frequency-dependent background signal, the characteristic timescale $\tau$ and the rms --velocity (for Equation \eqref{eq-bg-Harvey}) and intensity (for Equation \eqref{eq-bg-Karoff})-- $\sigma$ are different. For the two component background model, \citet{Jiang11} pointed out that the Karoff model can accurately represent the background signal in the frequency range of the oscillations. Here, we also adopt the Karoff model for the frequency-dependent background. Combined with the frequency independent white noise and the power excess hump from stellar oscillations, which is approximated by a Gaussian function, the overall power spectrum can be modeled by:
\begin{equation}\label{backg}
\begin{split}
P(\nu)=&P_{n} + \sum_{i=1}^{3} \frac{4\sigma_{i}^2\tau_{i}}{1+(2\pi\nu\tau_{i})^2+(2\pi\nu\tau_{i})^4} \\
&+ P_{g}~\mathrm{exp}~\left(\frac{-(\nu_{\mathrm{max}}-\nu)^2}{2\sigma_{g}^2}\right),
\end{split}
\end{equation}
where $P_{n}$ is the frequency independent white noise. For the Gaussian term, the parameters $P_{g}$, $\nu_{\mathrm{max}}$, and $\sigma_{g}$ denote the height, the central frequency, and the width of the power excess hump, respectively.

Similar to \citet{Jiang11}, we derive the pulsation spectrum through following four steps starting from the relative flux light curves: \rmnum{1}. compute the Fourier transformation to obtain the total power spectrum; \rmnum{2}. slightly smooth the total power spectrum using Gaussian smooth method with \textbf{F}ull \textbf{W}idth at \textbf{H}alf-\textbf{M}aximum (\textit{FWHM}) of 2 \muhz; \rmnum{3}. fit the total power spectrum, including the background signals and the oscillation signal, using Equation \eqref{backg} with $\sigma_{g}$ fixed to 1.28\dnu\ ($\textit{FHWM}=2\sqrt{2 \ln 2}\sigma_{g}\approx2.35\sigma_{g}$, hence $\sigma_{g}=1.28$\dnu\ corresponds to $\textit{FHWM}\approx3$\dnu) and obtain the central frequency \numax\ and its uncertainty (see panel (a) of Fig. \ref{power}); \rmnum{4}. remove the background from the power spectrum. From the above four steps, we obtain a clear oscillation power spectrum (see panel (b) of Fig. \ref{power}). We obtain the central frequency \numax, with a relative internal uncertainty of 0.6 percent except for three targets with much larger internal uncertainties ranging from 0.6 to 1.5 percents.

\subsubsection{Large frequency separation}\label{sec_deltanu}
The large frequency separation \dnu\ can be determined from the oscillation power spectrum using an \'{e}chelle diagram, i.e. this diagram shows oscillation frequencies vs. frequencies modulo \dnu, in which oscillations of the same degree form a near vertical ridge.
We apply the following steps to compute \dnu. (1). Compute an estimate of \dnu\ using the relation $\Delta\nu=\Delta\nu_{\odot}(\nu_{\mathrm{max}}/\nu_{\mathrm{max},\odot})^{0.784\pm0.003}$ \citep{hekker09aa}. (2). Change the value of \dnu\ in such a way that all oscillation modes of the same degree are located roughly on the same straight line with no systematic deviations (see panels (d-1)--(d-5) of Fig. \ref{power}). Or equivalently, all oscillation modes of the same degree overlap in frequency modulo \dnu\ (see panel (c) of Fig. \ref{power}).
Uncertainties in \dnu\ are obtained by applying small changes $\delta$\dnu\ to the value of \dnu\ until the modes have a systematic deviation in the \'{e}chelle diagram. We adopt the value of $\delta$\dnu\ as the uncertainty of \dnu. For the targets considered in the present work, the relative uncertainties are about 1.2\%. It should be noticed that such an estimation for the uncertainties of the averaged frequency separations is only a lower limit.

\begin{figure}
\begin{center}
  \includegraphics[scale=0.50,angle=-90]{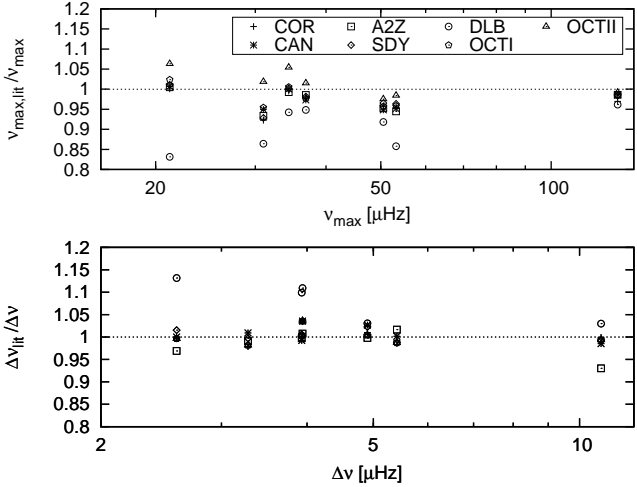}
  \caption{Relative values of literature values with respect to values obtained here vs. absolute values for \dnu\ (top panel) and \numax\ (bottom panel).  In the two panels, different symbols indicate results from different methods as indicated in the legend.
  The literature values are taken from Table~7. of \citet{Hekker11a}.}\label{fig-compare}
\end{center}
\end{figure}

In order to test the accuracy of our results, we compare our results with results from other methods \citep[including COR, CAN, A2Z, SDY, DLB, OCT\Rmnum{1}, and OCT\Rmnum{2},][]{Hekker11a} for a subset of stars (KIC 2424934, KIC 2424955, KIC 2425631, KIC 2444348, KIC 2448225, KIC 3526061, and KIC 3730953. This comparison is shown in Fig. \ref{fig-compare} and indicates that our results are in agreement with previous results.

\section{Masses RGB stars}\label{sec3}
\subsection{Scaling relations}\label{sec-classic}

From Equations \eqref{eq_deltanu} and \eqref{eq_numax}, the following relations can be obtained:
\begin{equation}\label{mass}
    \frac{M}{M_{\odot}}=\left(\frac{\Delta\nu}{\Delta\nu_{\odot}}\right)^{-4}\left(\frac{\nu_{\mathrm{max}}}{\nu_{\mathrm{max},\odot}}\right)^{3}\left(\frac{T_{\mathrm{eff}}}{T_{\mathrm{eff},\odot}}\right)^{3/2},
\end{equation}
\begin{equation}\label{radius}
    \frac{R}{R_{\odot}}=\left(\frac{\Delta\nu}{\Delta\nu_{\odot}}\right)^{-2}\left(\frac{\nu_{\mathrm{max}}}{\nu_{\mathrm{max},\odot}}\right)\left(\frac{T_{\mathrm{eff}}}{T_{\mathrm{eff},\odot}}\right)^{1/2},
\end{equation}
\begin{equation}\label{gravity}
    \frac{g}{g_{\odot}}=\left(\frac{\nu_{\mathrm{max}}}{\nu_{\mathrm{max},\odot}}\right)\left(\frac{T_{\mathrm{eff}}}{T_{\mathrm{eff},\odot}}\right)^{1/2}.
\end{equation}
The three relations are commonly used to calculate stellar mass $M$, radius $R$, and gravity $g$. The characteristics of these expressions were discussed by \citet{basu11}, who pointed out that there are larger uncertainties coming from the uncertainty in the effective temperatures. \citet{Miglio12} used Eqs.~\eqref{mass}-\eqref{gravity} to determine the parameters of RGB and RC stars in NGC 6791 and NGC 6819. The application to clump stars increases the uncertainties as a result of changes in stellar structure. For more detail information and discussions of the scaling relations and their uncertainties see \citet{White11,Miglio12,Mosser13,Belkacem13,hekker13}.

\begin{figure}
  \begin{center}
  \includegraphics[scale=0.58,angle=-90]{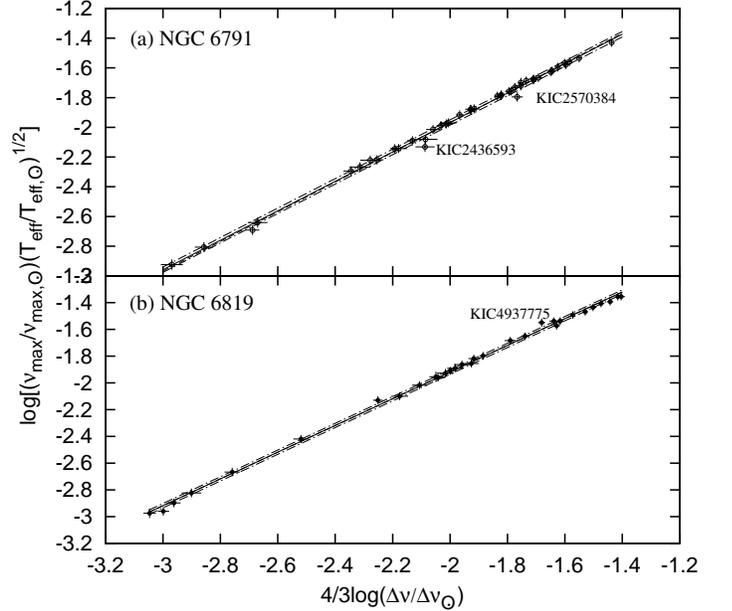}
  \caption{$\log[(\nu_{\mathrm{max}}/\nu_{\mathrm{max},\odot})\sqrt{T_{\mathrm{eff}}/T_{\mathrm{eff},\odot}}]~\rm{vs.}~4/3\log (\Delta\nu/\Delta\nu_{\odot})$. Panels (a) and (b) show results for NGC 6791 and NGC 6819, respectively. In the two panels, the solid lines show the fits (isomass lines) obtained with fitting relation (1) (Table~\ref{table_fit}) (free parameter fit); the dashed-dotted lines show the $1\sigma$ uncertainties of these fits; the long-dashed lines show the fits, with the parameter $A_{1}$ fixed to 1.0 (fixed parameter fit). The fitting parameters are listed in Table~\ref{table_fit}.}\label{fig_nu-t}
  \end{center}
\end{figure}

From Equations \eqref{mass}, we can obtain the following relation:
\begin{equation}\label{eq_num_log}
\log \left(\frac{\nu_{\mathrm{max}}}{\nu_{\mathrm{max},\odot}}\sqrt{\frac{T_{\mathrm{eff}}}{T_{\mathrm{eff},\odot}}}\right)=\frac{4}{3}\log \frac{\Delta\nu}{\Delta\nu_{\odot}}+\frac{1}{3}\log \frac{M}{M_{\odot}}.
\end{equation}
This shows that if the cluster stars have the same mass, then they should be located on a straight line when plotting $\log(\nu_{\rm{max}}/\nu_{\rm{max},\odot}\sqrt{T_{\rm{eff}}/T_{\rm{eff},\odot}})$ as the ordinate and $4/3\log(\Delta\nu/\Delta\nu_{\odot})$ as the abscissa (see Fig. \ref{fig_nu-t}).
The slope of the linear relationship is 1.0 and the intercept is $1/3\log(M/M_{\odot})$. Hence, stars with different masses would  be located in a series of parallel lines. The coefficients of fitting relation (1) in Table~\ref{table_fit} and Fig. \ref{fig_nu-t} show that Equation \eqref{eq_num_log} is in good agreement with the observations. Hence, we obtain the direct observational evidence for the masses of RGB stars in a cluster being almost a constant within their uncertainty (see also fitting relation (1) in Table~\ref{table_fit}). For the two considered clusters, the uncertainty is about 3\%.

From Fig. \ref{fig_nu-t} it is clear that  for NGC 6791 there are two stars (KIC 2436593 and KIC 2570384) that deviate from the straight line. For NGC 6819, all stars except for KIC 4937775 are located on a straight line.  For these three `outliers', we speculate that \teff\ is significantly different from the intrinsic value of \teff, which causes the deviations from the straight line.

Based on the above analysis, we obtain the following masses for RGB stars: $\bar{M}_{6791,\rm{I}}=1.25\pm0.03~M_{\odot}$ for NGC 6791 and $\bar{M}_{6819,\rm{I}}=1.75\pm0.05~M_{\odot}$ for NGC 6819. Compared to the recent results by \citet{Miglio12}  and \citet{basu11} --whom applied Equations \eqref{mass}-\eqref{gravity} and grid-based modeling, respectively, to the individual stars to obtain the mean RGB cluster masses--  our results are slightly higher (see Table~\ref{table_m}).

Irrespective of the method used, a significant uncertainty in the derived masses arises from the uncertainty in effective temperature \teff. Typically, the uncertainty of \teff\ is about 4\% ($\sim$150~K), which will lead to an uncertainty of $\sim$6\% in $M$ (see Table~\ref{table_uncertainty}). To decrease the uncertainties coming from the measurements of \teff, we should use more precise effective temperatures or use other parameters which can be determined with high precision to replace the effective temperature in the scaling relations. The latter solution is described in the next sections.

\renewcommand{\arraystretch}{1.3}
\begin{deluxetable}{ccc}
\tablewidth{0pt}
\tablecaption{Fitting relations and fitting coefficients.\label{table_fit}}
\startdata
\hline\hline
 (1) & \multicolumn{2}{c}{fit: $\log(\nu_{\mathrm{max}}\sqrt{T_{\mathrm{eff}}})=A_{1}\log\Delta\nu^{4/3}+B_{1}$} \\
     & \multicolumn{2}{c}{underlying Equation \eqref{eq_num_log}} \\
       ~~ &  $A_{1}$               &       $B_{1}$       \\% [0.3ex]
 \hline
 NGC 6791 & 0.991$\pm$0.009        &     0.01$\pm$0.02 \\
      ~   & 1.0\tablenotemark{a}    &     0.032$\pm$0.003 \\
 NGC 6819 & 0.998$\pm$0.008        &     0.08$\pm$0.02 \\
     ~    & 1.0\tablenotemark{a}    &     0.081$\pm$0.004 \\ %[0.3ex]
 \hline \hline
 (2) & \multicolumn{2}{c}{fit: $\log\sqrt{T_{\mathrm{eff}}}=A_{2}\log R+B_{2}$ } \\
     & \multicolumn{2}{c}{underlying Equation \eqref{eq_t}} \\
       ~~ &  $A_{2}$                 &       $B_{2}$        \\ %[0.3ex]
\hline
 NGC 6791 & $-$0.048$\pm$0.002     & $-$0.010$\pm$0.002 \\
   ~      & $-$0.049\tablenotemark{a} & $-$0.0094$\pm$0.0004 \\
 NGC 6819 & $-$0.049$\pm$0.003     & 0.002$\pm$0.003    \\
   ~      & $-$0.049\tablenotemark{a} & 0.0017$\pm$0.0007    \\
field stars& $-$0.041$\pm$0.001    & $-$0.006$\pm$0.001 \\
   ~      & $-$0.049\tablenotemark{a} & 0.0025$\pm$0.0003    \\ %[0.3ex]
 \hline \hline
(3) & \multicolumn{2}{c}{fit: $3\log\nu_{\mathrm{max}}=A_{3}\log\Delta\nu+B_{3}$ } \\
    & \multicolumn{2}{c}{underlying Equation \eqref{eq_new_log1}} \\
  ~~      &  $A_{3}$                 &       $B_{3}$        \\ %[0.3ex]
\hline
 NGC 6791 &  3.86$\pm$0.04       & 0.07$\pm$0.06 \\
    ~     &  3.902\tablenotemark{a}   & 0.13$\pm$0.01 \\
 NGC 6819 &  3.89$\pm$0.03       & 0.23$\pm$0.04 \\
 ~        &  3.902\tablenotemark{a}   & 0.25$\pm$0.01 \\
field stars& 3.92$\pm$0.01       & 0.13$\pm$0.02 \\
  ~       &  3.902\tablenotemark{a}   & 0.114$\pm$0.005
\enddata
\tablenotetext{a}{predictions}
\tablecomments{All the variables in expressed in solar units.}
\end{deluxetable}
\renewcommand{\arraystretch}{1}

\subsection{$\sqrt{T_{\mathrm{eff}}} \sim g^pR^q$}\label{sec-tgr}
In order to overcome the problem of a large uncertainty coming from the measurements of \teff, we replace \teff\ by other parameters in the scaling relations.

Stellar structure and evolution theory describes that, when a star exhausts its central hydrogen, it leaves the MS and evolves onto the RGB. In this process, the nuclear energy production takes place in a H-burning shell surrounding the inert He-core. In the RGB phase, the star is composed of a compact degenerate ($M\lesssim2.3$M$_{\odot}$) helium core and an extended convective envelope (about 99\% or more in radius). The properties of the RGB stars are similar to that of fully convective stars on the Hayashi track. In stellar structure and evolution theory, the Hayashi track stars can be approximated by a polytrop model. Using the polytrop model approximation, a relation $\sqrt{T_{\mathrm{eff}}} \sim g^{1/8}R^{5/16}$ or $\sqrt{T_{\mathrm{eff}}} \sim M^{1/8}R^{1/16}$ for stars on the Hayashi track can be established.

We suppose that for the RGB stars there is a similar relation, $\sqrt{T_{\mathrm{eff}}}\sim g^{p}R^{q}$, with different exponents ($p$ and $q$). This is because RGB stars have a degenerate helium core and a different composition in the stellar interior compared to stars on the Hayashi track.

We have calculated a series of stellar evolutionary tracks, covering a metallicity range from $-$0.8 to 0.4 dex with steps of 0.2 dex and using masses ranging from 0.8 to 2.0 $M_{\odot}$ with steps of 0.2 $M_{\odot}$. Analyzing these evolutionary tracks, taking into account the metallicity effect, we obtain a relation:
\begin{equation}\label{eq_t}
\sqrt{\frac{T_{\mathrm{eff}}}{T_{\mathrm{eff},\odot}}}=C(z)(\frac{R}{R_{\odot}})^{a}(\frac{M}{M_{\odot}})^{b},
\end{equation}
where $a=-0.0490\pm0.0002$, $b=0.051\pm0.002$, and
\begin{equation}\label{eq_t2}
C(z)=10^{c}(z/z_{\odot})^{d}
\end{equation}
with $c=-0.0080\pm0.0004$, $d=-0.0220\pm0.0005$ and $z_{\odot}=0.02$.
Hence,
\begin{equation}
\sqrt{T_{\mathrm{eff}}}\approx0.982z^{-0.022}R^{-0.049}M^{0.051}
\end{equation}
We ignore the small impact  from the stellar metallicity on the coefficients $a$ and $b$.
For the relation $\sqrt{T_{\mathrm{eff}}}\sim g^pR^q$, we find $p=0.051\pm0.002$ and $q=0.053\pm0.003$ for RGB stars.

\begin{figure}
  \begin{center}
  \includegraphics[scale=0.6,angle=-90]{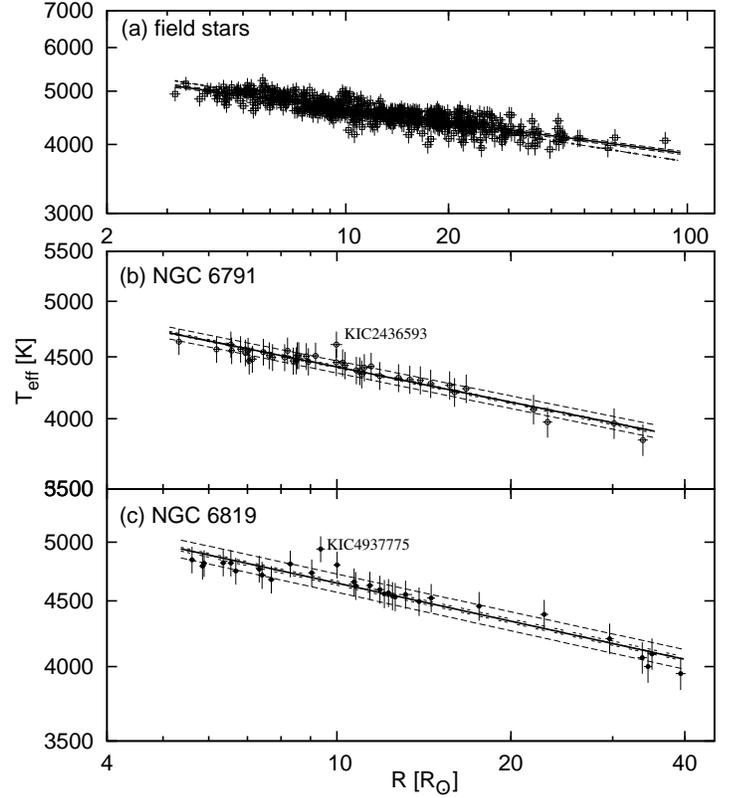}
  \caption{\teff\ vs. $R$. Panel (a) shows results for the field stars. Panels (b) and (c) show results for NGC 6791 and NGC 6819, respectively. The solid lines are the fits obtained with fitting relation $\log \sqrt{T_{\mathrm{eff}}}=A_{2}\log R+B_{2}$ (free parameter fit); the long-dashed lines show the corresponding 1$\sigma$ uncertainty; the dashed-dotted lines shows fits, with $A_{2}$ fixed to the predicted value of $-$0.049 (fixed parameter fit); and the dotted lines show the corresponding 1$\sigma$ uncertainty. The fit parameters are listed in Table~\ref{table_fit}. For field stars, the data points are taken from Table~7 of \citet{Hekker11a} (\dnu\ and \numax) and from \kic\ (\teff).}\label{fig1_t}
  \end{center}
\end{figure}

Note that the exponents $a$ and $b$, and the term $C(z)$ in Equation \eqref{eq_t} will depend on the mixing length parameter ($\alpha_{\mathrm{MLT}}$). However, the atmosphere model and the mixing length parameter $\alpha_{\mathrm{MLT}}=1.95$ used here, are suitable for both NGC 6791 and NGC 6819. This method leads to an intrinsic systematic dispersion of 80 K in \teff.

Observations provide sufficient evidence to prove the existence of Equation \eqref{eq_t}. Fig. \ref{fig1_t} shows that there is a clear correlation between the effective temperature and stellar radius, with a higher dispersion for field stars (panel (a)) compared to the two clusters (panel (b) and (c)). This dispersion is most likely caused by a larger range in metallicity and mass. For NGC 6791 all stars are consistent with the fitting relations as defined in Table~\ref{table_fit} except KIC 2436593. For NGC 6819 only KIC 4937775 deviates from the tight correlation. This could again be due to differences in effective temperature \teff.

For the clusters NGC 6791 and NGC6819, the fitting coefficients $A_{2}$ are $-0.048\pm0.002$ and $-0.049\pm0.003$, respectively. These are consistent with the predicted value of $-0.0490\pm0.0002$ (coefficient $a$). For the field stars, $A_{2}=-0.0408\pm0.0011$, which is slightly larger than the prediction.

Finally, combining Equations \eqref{eq_deltanu}, \eqref{eq_numax}, and \eqref{eq_t}, we obtain a relation between \teff, \dnu, \numax, and $z$:
\begin{equation}\label{eq_new_t}
(\frac{T_{\mathrm{eff}}}{T_{\mathrm{eff},\odot}})^{a+3b-1}=C(z)^{-2}(\frac{\Delta\nu}{\Delta\nu_{\odot}})^{4a+8b}(\frac{\nu_{\mathrm{max}}}{\nu_{\mathrm{max},\odot}})^{-2a-6b}.
\end{equation}
In this relation the effective temperature \teff\ is only related to the two oscillation parameters (\dnu\ and \numax) and a non-oscillation parameter ($z$). So, from parameters \dnu, \numax, and $z$, we could use this relation to estimate the stellar effective temperature \teff. Because the effective temperature is calculated from the oscillation parameters, we refer to it as the ``oscillation temperature" (\tosc).

\subsubsection{New expressions for $R$, $M$, and $g$ }\label{sec-new}

Combining Equations \eqref{eq_deltanu}, \eqref{eq_numax}, and \eqref{eq_t}, we obtain a series of new relations with respect to $R$, $M$, and $g$:
\begin{equation}\label{eq_new_r}
(\frac{R}{R_{\odot}})^{a+3b-1}=(\frac{\Delta\nu}{\Delta\nu_{\odot}})^{2-2b}(C(z)\frac{\nu_{\mathrm{max}}}{\nu_{\mathrm{max},\odot}})^{-1},
\end{equation}
\begin{equation}\label{eq_new_m}
(\frac{M}{M_{\odot}})^{a+3b-1}=(\frac{\Delta\nu}{\Delta\nu_{\odot}})^{4+2a}(C(z)\frac{\nu_{\mathrm{max}}}{\nu_{\mathrm{max},\odot}})^{-3},
\end{equation}
\begin{equation}\label{eq_new_g}
(\frac{g}{g_{\odot}})^{a+3b-1}=(\frac{\Delta\nu}{\Delta\nu_{\odot}})^{2a+4b}(C(z)\frac{\nu_{\mathrm{max}}}{\nu_{\mathrm{max},\odot}})^{-1}.
\end{equation}
A theoretical analysis shows that estimating stellar radius $R$, mass $M$, and gravity $g$ using these relations will lead to a systematic uncertainty of 1\% in $R$, of 2\% in $M$, and of 0.8\% in $g$, respectively.

\renewcommand{\arraystretch}{1.2}
\begin{deluxetable}{l|c|c|c|c|c|c}
\tablecaption{Intrinsic uncertainties of $M$, $R$, and $g$ for individual targets.\label{table_uncertainty}}
\tablewidth{0pt}
\startdata
\hline \hline
\multirow{2}{*}{Parameters} & \multicolumn{2}{c|}{$R$} & \multicolumn{2}{c|}{$M$} &\multicolumn{2}{c}{$g$} \\
\cline{2-7}
& Eq.~\ref{radius} & Eq.~\ref{eq_new_r} & Eq.~\ref{mass} & Eq.~\ref{eq_new_m} & Eq.~\ref{gravity} & Eq.~\ref{eq_new_g} \\
\hline
\dnu(1.2\%)  & 2.4\% & 2.5\% & 4.8\% & 5.2\% & \nodata & 0.1\% \\
\numax(1.5\%)& 1.5\% & 1.7\% & 4.5\% & 5.0\% & 1.5\%   & 1.7\% \\
\teff(4.0\%)\tablenotemark{a}  & 2.0\% &\nodata& 6.0\% &\nodata& 2.0\%   & \nodata\\
$z$(20.0\%)  &\nodata& 0.5\% &\nodata&1.5\%  & \nodata & 0.5\%  \\
\hline
total        & 3.5\% & 3.1\% & 8.9\% & 7.4\% & 2.5\%   & 1.8\%
\enddata
\tablecomments{Note that systematic uncertainties are not included in the total uncertainty.}
\tablenotetext{a}{For the two considered cluster, the uncertainty of \teff\ is about 3.0\%, but for field star the uncertainty of \teff\ usually larger than that. Here we fix it 4.0\% as a low limit.}
\end{deluxetable}
\renewcommand{\arraystretch}{1}

\begin{figure*}
\begin{center}
\includegraphics[scale=0.48,angle=-90]{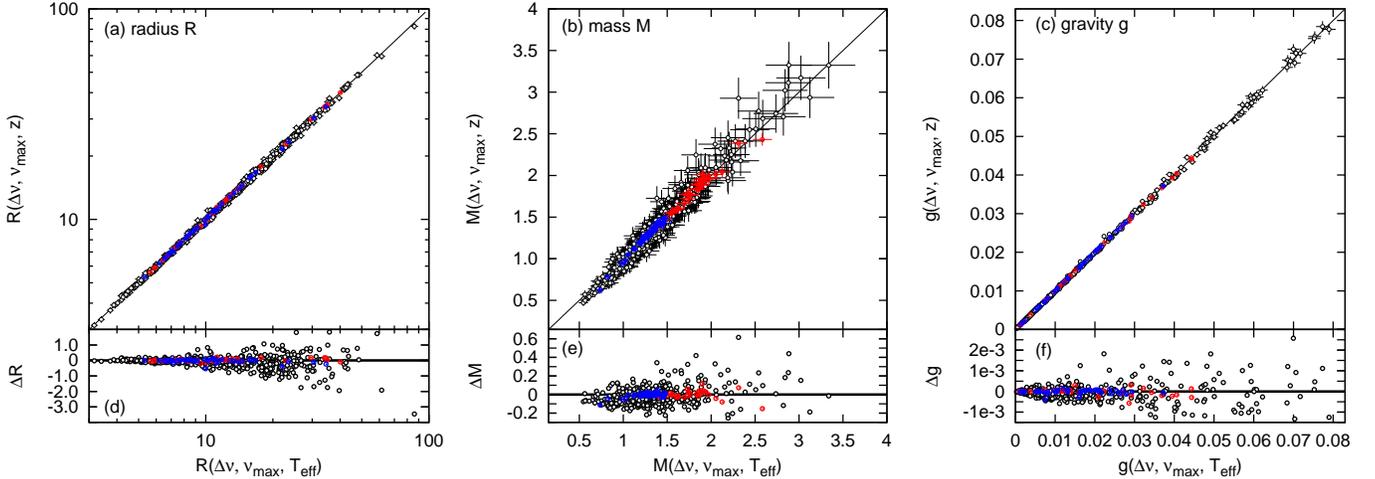}
\caption{Results from our method (Equations \eqref{eq_new_r} - \eqref{eq_new_g}) vs. previously used methods (Equations \eqref{mass}-\eqref{gravity}) to derive $R$, $M$, and $g$. The results for different parameters are shown in panels (a), (b), and (c), respectively, and their corresponding differences $\Delta R$, $\Delta M$, and $\Delta g$ between the two different methods are shown in panels (d), (e) and (f), respectively. In these panels, black open circles, blue dots and red dots represent the field stars, stars in NGC 6791 and NGC 6819, respectively. In panels (a), (b), and (c), diagonal thin solid lines are the isopleth lines, i.e. y=x. In panels (d), (e), and (f), thick solid lines represent y=0. }\label{fig_rmg}
\end{center}
\end{figure*}

Substituting the coefficients $a$, $b$, and $C(z)$ into Equations \eqref{eq_new_r}, \eqref{eq_new_m}, and \eqref{eq_new_g} and reorganizing them, we obtain the following relations:
\begin{equation}
R=0.980z^{-0.0246}\Delta\nu^{-2.118}\nu_{\mathrm{max}}^{1.116},
\end{equation}
\begin{equation}
M=0.940z^{-0.0737}\Delta\nu^{-4.355}\nu_{\mathrm{max}}^{3.348},
\end{equation}
\begin{equation}
 g=0.980z^{-0.0246}\Delta\nu^{-0.118}\nu_{\mathrm{max}}^{1.116},
\end{equation}
with all variables in solar units. The uncertainties in $M$, $R$ and $g$ due to propagation of uncertainties in the observables are listed in Table~\ref{table_uncertainty}. The main uncertainty in $R$, $M$ and $g$ derived from Equations \eqref{eq_new_r}-\eqref{eq_new_g} originate from the uncertainty of \dnu\ and \numax. In Equations \eqref{mass}-\eqref{gravity}, the uncertainties of the mass $M$ and gravity $g$ are mainly due to the uncertainty in \teff. The precision with which \dnu\ and \numax\ can be determined are higher than the precision with which \teff\ can be determined. Therefore, using Equations \eqref{eq_new_r}, \eqref{eq_new_m}, and \eqref{eq_new_g} to determine $R$, $M$, and $g$ can significantly reduce the uncertainties.

Combined with the system uncertainty, the total uncertainties of $R$, $M$, and $g$ derived from Equations \eqref{eq_new_r}-\eqref{eq_new_g} are about 4\%, 9.4\%, and 2.5\%, respectively. The systematic uncertainties from Equation \eqref{mass}-\eqref{gravity} are not derived.

In order to test Equations \eqref{eq_new_r}, \eqref{eq_new_m}, and \eqref{eq_new_g}, we compare the results with those of Equations \eqref{mass}-\eqref{gravity}. From the bottom panels of Fig. \ref{fig_rmg}, it can be seen that for the two methods, both the stellar radius and gravity are in good agreement, while the stellar mass has larger dispersion. On the whole, there is good agreement, especially for the cluster stars.

\subsection{\dnu\ - \numax\ relation}
From Equation \eqref{eq_new_m}, we can obtain the following relation between \numax\, \dnu\, $M$ and $z$:
\begin{equation}\label{eq_new_log}
\begin{split}
3\log (\frac{\nu_{\mathrm{max}}}{\nu_{\mathrm{max},\odot}})-&(4+2a)\log (\frac{\Delta\nu}{\Delta\nu_{\odot}})= \\
&(1-3b-a)\log (\frac{M}{M_{\odot}})-3\log C(z).
\end{split}
\end{equation}
Substituting the coefficients $a$, $b$, and $C(z)$ this can be rewritten as:
\begin{equation}\label{eq_new_log1}
\begin{split}
3\log & (\frac{\nu_{\mathrm{max}}}{\nu_{\mathrm{max},\odot}})-3.902\log (\frac{\Delta\nu}{\Delta\nu_{\odot}})= \\
& 0.896\log (\frac{M}{M_{\odot}})+0.066\log (\frac{z}{z_{\odot}})+0.024.
\end{split}
\end{equation}
This equation explicitly expresses the relation between \numax, \dnu, $M$, and $z$. There are similar relations between \numax\ and \dnu\ established from observations. These are expressed as $\Delta\nu\approx\alpha(\nu_{\mathrm{max}}/\mu\rm{Hz})^{\beta}$ with values for $\alpha$ in the range $[0.254\pm 0.004,0.293 \pm 0.009]  \mu$Hz and $\beta$ in the range $[0.745\pm0.003,0.772\pm0.005]$ \citep[see][and references therein]{huber10}, or as $\Delta\nu=\Delta\nu_{\odot}(\nu_{\mathrm{max}}/\nu_{\mathrm{max},\odot})^{\beta}$ with $\beta = 0.784\pm0.003$ \citep{hekker09aa}.

We can express Equation \eqref{eq_new_log1} in a similar form as:
\begin{equation}\label{dnu_numax}
\begin{split}
\Delta\nu=0.986\Delta\nu_{\odot}&(\nu_{\mathrm{max}}/\nu_{\mathrm{max},\odot})^{0.769}\\
&(M/M_{\odot})^{-0.230}(z/z_{\odot})^{-0.017}.
\end{split}
\end{equation}
In this predicted relation, the exponent $\beta=0.769$ is consistent with previous observational results. At the same time, we express the parameter $\alpha$ as a function of $M$ and $z$. The relation derived here is based on the structure and evolution of RGB stars, and thus provides additional information about RGB stars and RGB stars only. Contrary to the previously derived relations, which are applicable to both main sequence and red-giant stars (RGB \& RC).

\begin{figure}
\begin{center}
\includegraphics[scale=0.58,angle=-90]{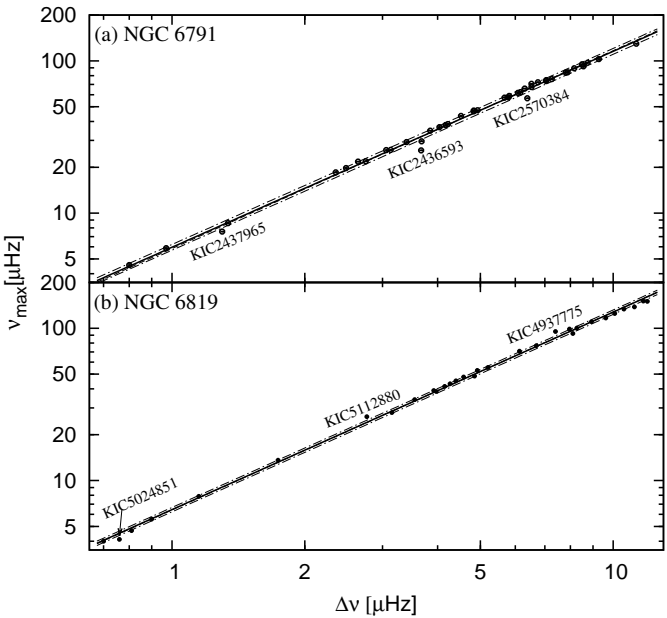}
\caption{\numax\ vs. \dnu. Panels (a) and (b) show \numax\ vs. \dnu\ for NGC 6791 and NGC 6819, respectively. The solid lines represent the fits following the relation $3\log\nu_{\mathrm{max}}=A_{3}\log\Delta\nu+B_{3}$ (free parameter fit); the dashed-dotted lines represent the corresponding  $1\sigma$ uncertainty; the short-dashed lines represent the fits, with $A_{3}$ fixed to the predicted value of 3.902 (fixed parameter fit). The values of $A_{3}$ and $B_{3}$ are listed in Table~\ref{table_fit}.}\label{fig-delta-nu_nu}
\end{center}
\end{figure}

For the observations, we fit \numax\ vs. \dnu\ with the following relation
\begin{equation}
3\log\nu_{\mathrm{max}}=A_{3}\log\Delta\nu+B_{3},
\end{equation}
with all variables in solar units. For $A_{3}$ we find 3.9 for both the cluster stars and the field stars  consistent with the predicted value of 3.902. The results are shown in Fig. \ref{fig-delta-nu_nu} and Table~\ref{table_fit}. All except six stars --which are indicated in the figure-- follow a tight relation, i.e. these six stars have different determined masses compared to the other target stars. The three stars in NGC 6791  have masses below the masses of the other stars, while in NGC 6819, there are two stars with larger and one star with lower mass than the majority of stars.

Equation \eqref{eq_new_log1} and fitting relation (3) (Table~\ref{table_fit}) show that the fitting coefficient $B_{3}$ is related to both stellar mass $M$ and metallicity $z$. Hence, when one parameter is constraint, we can derive the other.
In Sect. \ref{sec-iso}, we obtained the cluster metallicities $z_{6791,\rm{iso}}=0.040\pm0.005$ and $z_{6819,\rm{iso}}=0.022\pm0.004$ from isochrones. These metallicities combined with Equation \eqref{eq_new_log1} and fitting relation (3) of Table~\ref{table_fit} provides the masses of RGB stars in the two clusters. The masses are $\bar{M}_{6791,\rm{II}}=1.24\pm0.03M_{\odot}$ and $\bar{M}_{\rm 6819,\rm{II}}=1.77\pm0.05M_{\odot}$ for NGC 6791 and NGC 6819, respectively.

Additionally from Equation \eqref{eq_t} and fitting relation (2) (Table~\ref{table_fit}), we obtain a relation between the fitting coefficient $B_{2}$, mass $M$, and metallicity $z$:
\begin{equation}\label{eq-B2-mz}
\log C(z)+b\log M=B_{2}.
\end{equation}
Combining Equation \eqref{eq_new_log} and fitting relation (3) (Table~\ref{table_fit}), we can establish another relation:
\begin{equation}\label{eq-B3-mz}
-3\log C(z)+(1-3b-a)\log M=B_{3}.
\end{equation}
Substituting the fitting coefficients $B_{2}$, $B_{3}$ and theoretical model coefficients $a$, $b$, and $C(z)$ into these equations (Equations \eqref{eq-B2-mz} and \eqref{eq-B3-mz}) we find the masses and metallicties simultaneously. They are $\bar{M}_{6791,\rm{III}}=1.25\pm0.03M_{\odot}$ and $z_{6791,\rm{III}}=0.039\pm0.002$ for NGC 6791 and $\bar{M}_{6819,\rm{III}}=1.75\pm0.04 M_{\odot}$ and $z_{6819,\rm{III}}=0.026\pm0.002$ for NGC 6819.

\section{Results and discussions}\label{sec4}

\subsection{$R$, $M$, and $g$ of individual targets}

Stellar mass $M$, radius $R$, and gravity $g$ are important basic parameters in stellar physics. Accurate values for these parameters can be obtained from the gravitational interaction between stars in multiple systems or from asteroseismology. In asteroseismology, Equations \eqref{mass}-\eqref{gravity} are usually used to estimate $M$, $R$, and $g$ from the three observables \dnu, \numax, and \teff. In those relations, the uncertainty of effective temperature \teff\ introduces the largest uncertainty. For an individual RGB star, the uncertainty in effective temperature is around $\sim$150 K ($\sim$4\%). It leads to an uncertainty of $\sim$6\% in mass, of $\sim$2\% in radius, and of $\sim$2\% in gravity (for detail see Table~\ref{table_uncertainty}).

Using $T_{\rm osc}$ (Equation \eqref{eq_new_t}) in stead of \teff\ in Equations \eqref{mass}-\eqref{gravity}, we have obtained a new series of relations (Equations \eqref{eq_new_r}, \eqref{eq_new_m}, and \eqref{eq_new_g}) to derive $M$, $R$ and $g$ from \dnu, \numax, and $z$. Due to the smaller weight $w_{z}$  of $z$ in Equation \eqref{eq_new_t} compared to the weights of \dnu\ and \numax\ ($w_{\nu_{\mathrm{max}}}/w_{z}\approx w_{\Delta\nu}/w_{z}\approx5$), the uncertainties in $M$, $R$ and $g$ from Equations \eqref{eq_new_r}, \eqref{eq_new_m}, and \eqref{eq_new_g} are mainly dominated by uncertainties in \dnu\ and \numax, which can be measured with much higher precision than \teff. Usually, the intrinsic uncertainties in \dnu\ and \numax\ are below 1.5\%.

The results from the two methods described above are consistent, except for a few outliers. The field stars have larger dispersions compared to cluster stars, which we attribute to the larger variation in \teff\ and $z$ in field stars.

From the above analysis, we suggest that using Equations \eqref{eq_new_r}, \eqref{eq_new_m}, and \eqref{eq_new_g} to estimate $R$, $M$, and $g$ is a better choice.

\subsection{Cluster mass and metallicity}

To determine the masses of RGB stars in clusters, \citet{basu11} used so-called grid-based modeling. Under the condition that the masses of RGB stars in a cluster are constant, they obtained a mean mass value. In \citet{Miglio12}, masses for the individual stars are derived from Equations \eqref{mass}-\eqref{gravity} and an average mass is computed similar to \citet{basu11}. In the present study, we do not calculate individual stellar masses for these RGB stars in the clusters, but regard all RGB stars in a cluster as one entity, and calculate the average mass of its members.

\begin{deluxetable}{cllll}
\tablecaption{Mass and metallicity of RGB stars in NGC 6791 and NGC 6819.\label{table_m}}
\tablewidth{0pt}
\tablehead{
\colhead{} & \colhead{NGC 6791} & \colhead{} & \colhead{NGC 6819} & \colhead{} \\
\colhead{Methods} & \colhead{$\bar{M}(M_{\odot})$} & \colhead{$z$ or [Fe/H]} & \colhead{$\bar{M}(M_{\odot})$} & \colhead{$z$ or [Fe/H]}
}
\startdata
 (I)\tablenotemark{a}   & 1.25$\pm$0.03 & $0.040\pm0.005$\tablenotemark{b,c} & 1.75$\pm$0.05 & $0.022\pm0.004$\tablenotemark{b,c} \\
 (II)\tablenotemark{a}  & 1.24$\pm$0.03 & $0.040\pm0.005$\tablenotemark{b,c} & 1.77$\pm$0.05 & $0.022\pm0.004$\tablenotemark{b,c} \\
 (III)\tablenotemark{a} & 1.25$\pm$0.03 & $0.039\pm0.002$ & 1.75$\pm$0.04 & $0.026\pm0.002$ \\
 (IV)\tablenotemark{d}  & 1.20$\pm$0.01 & $\text{[Fe/H]}=+0.3$\tablenotemark{c} & 1.68$\pm$0.03 & $\text{[Fe/H]}=0.0$\tablenotemark{c} \\
 (V)\tablenotemark{e}   & 1.23$\pm$0.02 & $\text{[Fe/H]}=+0.3$\tablenotemark{c} & 1.61$\pm$0.04 & $\text{[Fe/H]}=0.0$\tablenotemark{c}
\enddata
\tablenotetext{a}{The present work.}
\tablenotetext{b}{Derived from fitting isochrone.}
\tablenotetext{c}{As a input parameters to determine \teff\ and/or further to determine the mass.}
\tablenotetext{d}{\citet{basu11}.}
\tablenotetext{e}{\citet{Miglio12}.}
\end{deluxetable}

\subsubsection{NGC 6791}

For NGC 6791, the previously obtained average mass values of the RGB stars are $1.20\pm0.01~M_{\odot}$ \citep{basu11} and $1.23\pm0.02~M_{\odot}$ \citep{Miglio12}. In the present study, we have obtained three mean masses for the RGB stars of the cluster with different methods. They are $\bar{M}_{6791,\rm{I}}=1.25\pm0.03~M_{\odot}$, $\bar{M}_{6791,\rm{II}}=1.24\pm0.03~M_{\odot}$, and $\bar{M}_{6791,\rm{III}}=1.25\pm0.02~M_{\odot}$. These are listed in Table~\ref{table_m}.

\subsubsection*{$M_{\rm{I}}$}

$\bar{M}_{\rm{I}}$ has been derived from Equation \eqref{eq_num_log}. We have applied fit 1 (see Table~\ref{table_fit}) to the data (\dnu, \numax, and \teff) and obtained the average  RGB cluster mass from combining coefficients $A_{1}$ and $B_{1}$  and Equation \eqref{eq_num_log}. This resulted in $\bar{M}_{6791,\rm{I}}=1.25\pm0.03~M_{\odot}$. Panel (a) of Fig. \ref{fig_nu-t} shows that all targets (except for KIC 2436593 and KIC 2570384) are consistent with the fit. The two outliers could be caused by: (1) biases in \teff; (2) genuinly different (smaller) masses. Nevertheless, Equation \eqref{eq_num_log} and Fig. \ref{fig_nu-t} show that the masses of RGB stars in a cluster can be regarded as a constant within a certain uncertainty.

\subsubsection*{$M_{\rm{II}}$}

$\bar{M}_{\rm{II}}$ is derived from Equation \eqref{eq_new_log}. This equation expresses the relationship among the four variables \dnu, \numax, $M$, and $z$. Fig. \ref{fig_nu-t} and Table~\ref{table_fit} show that the prediction is consistent with fits to the observed data. The fitting results can be expressed as $3\log\nu_{\mathrm{max}}\approx3.9\log\Delta\nu+B_{3}$. Combining with Equation \eqref{eq_new_log} it can be shown that $B_{3}$ is related to the masses and metallicity of the RGB stars. A similar relation $\Delta\nu=\alpha\cdot\nu_{\rm max}^{\beta}$ has been obtained from observational data. However, slightly different values for the coefficients $\alpha$ and $\beta$ have been derived \citep[e.g.][]{hekker09aa,stello09mn,Mosser10,huber10,Hekker11b,Hekker11c}. The differences in $\alpha$ and $\beta$ are due to the mass dispersion of this relation (see the bottom left panel of Fig.~5 in \citet{Hekker11b}), which is not incorporated in the coefficients.
Here, we provide an explicit physical meaning for the coefficient $\alpha$, i.e. we express the coefficient $\alpha$ as a function of $M$ and $z$ ($\alpha=0.986\Delta\nu_{\odot}(\nu_{\mathrm{max},\odot}/\mu\rm{Hz})^{-0.769}(M/M_{\odot})^{-0.230}(z/z_{\odot})^{-0.017}$). Thus, the coefficient $\alpha$ is inversely correlated with both $M$ (predominantly) and $z$. This is consistent with the results of \citet{Hekker11b}. From this relation, we can estimate the masses or metallicities of the RGB stars in the clusters.

Three stars, KIC 2436593, KIC 2437965, and KIC 2570384, are not consistent with the prediction (see panel (a) of Fig. \ref{fig-delta-nu_nu}). The deviations from the predictions are most likely due to the stellar masses, as the weight of mass is much larger than that of the metallicity in Equation \eqref{eq_new_log1}. Hence these targets have smaller determined masses compared to the other targets in the cluster. This is consistent with the results of $M_{\rm{I}}$.
From Equation \eqref{eq_new_log1} and using $z_{6791,\rm{iso}}=0.040\pm0.005$ estimated from isochrones \citep[in good agreement with $z\approx0.04\pm0.01$][]{Hekker11b}, we have obtained $\bar{M}_{6791,\rm{II}}=1.24\pm0.03~M_{\odot}$.

\subsubsection*{$M_{\rm{III}}$}

Equations \eqref{eq_t} and \eqref{eq_new_log} are both related to $M$ and $z$. Therefore, both parameters can be obtained when combining the two equations. Thanks to the near constant mass and metallicity of RGB stars of a cluster it is indeed possible to combine Equations \eqref{eq_t} and \eqref{eq_new_log} and fitting relations (1) and (3) of Table~\ref{table_fit} to simultaneously obtain $\bar{M}_{6791,\rm{III}}=1.25\pm0.03~M_{\odot}$ and $z_{6791,\rm{III}}=0.039\pm0.002$.

The metallicity $z_{6791,\rm{III}}=0.039\pm0.002$, it is consistent with $z_{6791,\rm{iso}}=0.04\pm0.005$ and with \citet{Hekker11b} $z\approx0.04\pm0.01$. Also for the mass $\bar{M}_{6791,\rm{III}}=1.25\pm0.03~M_{\odot}$ is consistent with $\bar{M}_{6791,\rm{I}}=1.25\pm0.03~M_{\odot}$ and $\bar{M}_{6791,\rm{II}}=1.24\pm0.03~M_{\odot}$. Therefore, the mass and the metallicity are self-consistent for the cluster when using these methods.

Finally, our results are consistent with \citet{Miglio12} $1.23\pm0.02~M_{\odot}$, but slightly larger than $1.20\pm0.01M_{\odot}$ obtained by \citet{basu11}, although still well within 3$\sigma$. From the isochrones we find a lower mass of $1.15\pm0.01M_{\odot}$. The cluster metallicity, $z_{6791}=0.039\pm0.002$ is consistent with \citet{basu11} and \citet{Miglio12} who find $0.04\pm0.01$, and with the value $0.04\pm0.005$, which comes from fitting isochrones in Sect. \ref{sec-iso}, and also in good agreement with \citet[][]{Brogaard11} who obtain $\rm{[Fe/H]}=+0.29\pm0.10$ dex from the spectroscopic observations.
Furthermore, we computed an age using a stellar model with the asteroseismic results as input. We find $6.0\pm0.5$ Gyr, which is considerably younger than the isochrone age of $8.0\pm0.4$ Gyr and also younger than ages commonly mentioned in the literature which ranges from 7 to 12 Gyr \citep[see e.g.][]{basu11,grundahl08}.

\subsubsection{NGC 6819}
For NGC 6819, the previously obtained average mass values of the RGB stars are $1.68\pm0.03~M_{\odot}$ \citep{basu11} and $1.61\pm0.04~M_{\odot}$ \citep{Miglio12}. In the present study, we have obtained three mean masses for the RGB stars of the cluster with different methods. They are $\bar{M}_{6819,\rm{I}}=1.75\pm0.05~M_{\odot}$, $\bar{M}_{6819,\rm{II}}=1.77\pm0.05~M_{\odot}$, and $\bar{M}_{6819,\rm{III}}=1.76\pm0.03~M_{\odot}$. These are listed in Table~\ref{table_m}.

From the three results for NGC 6819, $\bar{M}_{6819,\rm{II}}=1.77\pm0.05M_{\odot}$ is a slightly higher (although consistent) compared to $\bar{M}_{6819,\rm{III}}=1.75\pm0.03M_{\odot}$. This is caused by a difference in cluster metallicity. To derive $M_{\rm{II}}$ we used $z_{6819,\rm{iso}}=0.022\pm0.004$) derived from isochrones, while in method III we find $z_{6819,\rm{III}}=0.026\pm0.002$. If we substitute the metallicity $z_{6819,\rm{III}}=0.026\pm0.002$ into method II we obtain $\bar{M}_{6819,\rm{II}}^{'}=1.75\pm0.03M_{\odot}$ in agreement with the results from methods I and III. For self-consistency, we propose the cluster metallicity to be $z=0.026\pm0.002$.

Compared to other results in the literature, our mass values are larger than the previously obtained values of $1.68\pm0.03M_{\odot}$ \citep{basu11}, and $1.61\pm0.04M_{\odot}$ \citet{Miglio12}. This discrepancy could come from \teff, which is determined with $(B-V)$, and the value of [Fe/H]. These are  slightly higher than the values obtained from $(V-K)$, which are used by \citet{basu11} and \citet{Miglio12}. For the metallicity, our result $z=0.026\pm0.002$ is consistent with \citet[][]{Bragaglia01} who obtain $\rm{[Fe/H]}=+0.09\pm0.03$ dex from the high-dispersion spectroscopy.

From isochrone fitting we obtained $\bar{M}_{6819,\rm{iso}}=1.73\pm0.02M_{\odot}$ and $z_{6819,\rm{iso}}=0.022\pm0.004$. Comparing these results with those derived from asteroseismology, we find that the mass and the cluster metallicity are consistent with each other. Additionally, we computed an age using a stellar model with the asteroseismic results ($\bar{M}_{6819}=1.75\pm0.03M_{\odot}$ and $z_{6819}=0.026\pm0.002$). In this way we obtain an age of $1.8\pm0.1$ Gyr. This is consistent with the isochrone age of $1.9\pm0.1$ Gyr, but lower than the literature age of about 2.5~Gyr \citep{kalirai01,kalirai04}.

\subsection{Method comparison}

There are three methods used to determine the masses of RGB stars for cluster NGC 6791 and NGC 6819. For the three methods, the functions are and therefore each method has a different sensitivity to uncertainties in the observables. The source and effects of measurement uncertainties are listed in Table~\ref{table_uncertainty_method}. The additional information on the functions is listed in Table~\ref{table_comparison-method}.

\renewcommand{\arraystretch}{1.2}
\begin{deluxetable}{l|c|c|c}
\tablecaption{Source and effects of measurement or derived uncertainties.\label{table_uncertainty_method}}
\tablewidth{0pt}
\startdata
\hline \hline
\multirow{2}{*}{Parameters}  & Method I & Method II & Method III \\
                             & (Eq. \eqref{eq_num_log})& (Eq. \eqref{eq_new_log1}) & (Eqs \eqref{eq_t} and \eqref{eq_new_log1})\\
\hline
\dnu(1.2\%)  & 4.8\% & 5.2\% & 4.5\%  \\
\numax(1.5\%)& 4.5\% & 5.0\% & 4.3\%   \\
\teff(3.0\%) & 4.5\% &\nodata& 4.3\%   \\
$z$(20.0\%)  &\nodata& 1.5\% &\nodata \\
$R$(3.2\%)\tablenotemark{a}   &\nodata&\nodata& 0.5\%  \\
\hline
total        & 8.0\% & 7.4\% & 7.6\%
\enddata
\tablenotetext{a}{Derived from Equation \eqref{radius}.}
\end{deluxetable}
\renewcommand{\arraystretch}{1}

\renewcommand{\arraystretch}{1.3}
\begin{deluxetable}{c p{7cm}}
\tablecaption{Additional function information.\label{table_comparison-method} }
\tablewidth{0pt}
\tablehead{
\colhead{Methods} & \colhead{Functions}
}
\startdata
 (I)
 & Provides direct evidence that the masses of RGB stars in a cluster are the same within their uncertainty (See Fig. \ref{fig_nu-t}).
 \\
 (II)
 & Provides explicit and clear physical meaning and accurate expression for the coefficient $\alpha$ in the relation $\Delta\nu\approx\alpha(\nu_{\mathrm{max}}/\mu\rm{Hz})^{\beta}$, i.e. the coefficient $\alpha$ is expressed as a function of $M$ and $z$.
 \\
 (III)
 & Provides both $M$, and $z$ in a self consistent way.
\enddata
\end{deluxetable}
\renewcommand{\arraystretch}{1.0}

\subsection{KIC 2436593 and KIC 4937775}
In Figs \ref{fig_nu-t}, \ref{fig1_t}, and \ref{fig-delta-nu_nu}, there are a few outliers. Their properties, most likely masses and/or temperatures, differ from the other targets. For KIC 2436593 our analysis and the results by \citet{Stello11b}  suggest that this is a blended target. KIC 4937775 is a member of a binary system \citep{Hole09}. Therefore, it seems likely that (1) the color-index is influenced by the companion star, or (2) there might be material exchange in the binary system. Although the second option is more speculative.

\section{Summary and conclusions}\label{sec5}
From the global oscillation parameters (large frequency separation \dnu\ and frequency of maximum oscillation power \numax) and effective temperature \teff, we have determined the masses and metallicities of RGB stars in the clusters  NGC 6791 and NGC 6819, using newly devised relations.  From this investigation we conclude the following:

\rmnum{1}: Our method provides direct observational evidence that the masses of RGB stars in a cluster are the same within their uncertainty. In addition, we have determined their masses to be $1.24\pm0.03M_{\odot}$ and $1.75\pm0.05M_{\odot}$ for NGC 6791 and NGC 6819, respectively.

\rmnum{2}: Using the relation $\sqrt{T_{\rm eff}} \sim g^pR^q$ for stars on the Hayashi track calibrated with a grid of models, we have obtained a relation $\sqrt{T_{\mathrm{eff}}}\approx0.9820z^{-0.0220}R^{-0.0490}M^{0.0510}$ between \teff, $R$, $M$, and $z$ for red-giant stars. This relation has been verified by observations.

\rmnum{3}: Based on the above effective temperature relation, a series of relations with respect to $R$, $M$, and $g$ have been obtained. They can be used to estimate the stellar radius $R$, mass $M$, and surface gravity $g$ from oscillation parameters (\dnu\ and \numax) and metallicity $z$. Their uncertainties mainly come from uncertainties of the oscillation parameters (\dnu\ and \numax). The uncertainty of $z$ only slightly influence these relations.

\rmnum{4}: From analysis of models, we have obtained a relation $3\log\nu_{\mathrm{max}}-3.902\log\Delta\nu=0.896\log M+0.066\log z+0.024$, which accurately represent the relationship between \dnu, \numax, $M$, and $z$. We have verified this relation using observational data and derived masses and metallicities of the RGB stars in the two considered clusters. For NGC 6791 we find $\bar{M}_{6791}=1.25\pm0.02M_{\odot}$ and $z_{6791}=0.039\pm0.002$. For NGC 6819 we find $\bar{M}_{6819}=1.75\pm0.03M_{\odot}$ and $z_{6819}=0.026\pm0.002$.

Including these asteroseismic results in stellar models, we can estimate the cluster ages. They are about $6.0\pm0.5$ Gyr for NGC 6791 and about $1.8\pm0.1$ Gyr for NGC 6819. Our result for NGC 6819 is consistent with the result of fitting an isochrone ($1.85\pm0.1$ Gyr). For NGC 6791 the obtained age using a stellar model with asteroseismic input is lower than the result of fitting an isochrone ($8.0\pm0.4$ Gyr).

\acknowledgments
This work is co-sponsored by the NSFC of China (Grant Nos. 11333006 and 10973035), and by the Chinese Academy of Sciences (Grant No. KJCX2-YW-T24). The authors express their sincere thanks to NASA and the \kepler\ team for allowing them to work with and analyze the \kepler\ data making this work possible. The \keplermission\ is funded by NASA's Science Mission Directorate. The authors also gratefully acknowledge the computing time granted by the Yunnan Observatories, and provided on the facilities at the Yunnan Observatories Supercomputing Platform. SH acknowledges support from the European Research Council under the European Community's Seventh Framewrok Programme (FP7/2007-2013) / ERC grant agreement no 338251 (StellarAges). Fruitful discussions with C. Y. Ding, X. S. Fang, X. J. Lai, J. Su, Y. B. Wang, and Q. S. Zhang are highly appreciated. In addition, the authors want to pay tribute (heart-felt thanks) to K. Brogaard, B. Mosser, and D. Stello for advise through the \textit{KASC} (\kepler\ Asteroseisic Science Center) review. The authors also express their sincere thanks to Prof. Achim Weiss for productive advices. The authors are cordially grateful to an anonymous referee for his/her instructive advice and productive suggestions.

\end{document}